# Spin Hall Effect in Bilayer Graphene Combined with an Insulator up to Room Temperature

C. K. Safeer[1], Josep Ingla-Aynés[1], Nerea Ontoso[1], Franz Herling[1], Wenjing Yan[1,2], Luis E. Hueso[1,3], Fèlix Casanova[1,3,*].

[1] CIC nanoGUNE BRTA, 20018 Donostia-San Sebastian, Basque Country, Spain.
[2] School of Physics & Astronomy, The University of Nottingham, Nottingham NG7 2RD, UK
[3] IKERBASQUE, Basque Foundation for Science, 48013 Bilbao, Basque Country, Spain.

*E-mail: f.casanova@nanogune.eu

**Abstract:** Spin-orbit coupling in graphene can be enhanced by chemical functionalization, adatom decoration or proximity with a van der Waals material. As it is expected that such enhancement gives rise to a sizeable spin Hall effect, a spin-to-charge current conversion phenomenon of technological relevance, it has sparked wide research interest. However, it has only been measured in graphene/transition metal dichalcogenide van der Waals heterostructures with limited scalability. Here, we experimentally demonstrate spin Hall effect up to room temperature in bilayer graphene combined with a nonmagnetic insulator, an evaporated bismuth oxide layer. The measured spin Hall effect raises most likely from an extrinsic mechanism. With a large spin-to-charge conversion efficiency, scalability, and ease of integration to electronic devices, we show a promising material heterostructure suitable for spin-based device applications.

Spin-orbit coupling (SOC) is a basic ingredient in condensed matter that leads to many different phenomena in magnetism and spintronics[1–3]. Some of the most studied SOC-induced phenomena are the (inverse) spin Hall effect [(I)SHE][4] and the (inverse) Rashba-Edelstein effect [(I)REE)][5] due to their technological applicability. These effects enable the generation and detection of spin currents, which are being used to develop non-volatile memories[6–8] and spin-based logic devices[9,10]. Whereas materials with heavy atoms possess strong SOC, graphene, a two-dimensional material with extraordinary properties, is made of light carbon atoms and possesses a weak intrinsic SOC[11]. Therefore, it was considered as an unfavorable candidate to build active spintronic devices. However, several studies suggest that a significant SOC can be induced in graphene via chemical functionalization[12,13], heavy metal adatom decoration[14–18] or due to a proximity effect from a neighboring van der Waals material[19–22]. The latter case has been proven to be very efficient, especially by combining graphene with a transition metal dichalcogenide (TMD), a two-dimensional material family with strong SOC[23]. The creation of such van der Waals heterostructures induces an enhancement of SOC in the order of few meV[19–22], which has been demonstrated to lead to weak antilocalization[24–26] and spin lifetime anisotropy[27–29]. Furthermore, the gate tunability of spin absorption by the TMD in such systems paves the way for active control of spin currents in graphene/TMD heterostructures[30,31].

Steaming from this enhancement of SOC, theoretical studies have shown that SHE in graphene can be realized due to a modification of the band structure via proximity effect or when graphene is decorated with adatoms, molecules, or nanoparticles. The former case is predicted to occur in graphene/TMD heterostructures, where the SHE is dominated by a uniform valley-Zeeman SOC which arises due to the breaking of the sublattice symmetry (intrinsic mechanism)[22,32,33]. In the latter case, some parts of the graphene sample remain intact, but large clusters of impurities generate spatially fluctuating spin-orbit fields which are an extrinsic source for SHE[14–17]. Experimentally, the proximity-induced SHE has been the only one demonstrated in graphene, showing large spin-charge interconversion efficiencies[34–36].

Along with the understanding of the mechanisms behind the SHE in graphene, it is also important to explore graphene-based spin-to-charge conversion systems suitable for device applications. In this respect, graphene/TMD van der Waals heterostructures have two inherent problems. Firstly, although the fabrication process of using mechanical exfoliation followed by deterministic transfer techniques is very successful in a research environment,[34–41] it has limited scaling capabilities. Therefore, obtaining SHE in graphene combined with heavy metal-based materials that can be grown using standard deposition tools would be more suitable for scalable device fabrication. Secondly, in all reported experiments, graphene has been combined with metallic or semiconducting materials that usually shunt the spin and charge currents. This reduces the accumulated charge or spin density in graphene, which is detrimental to device applications in which large spin or charge current densities are required. This issue might be solved by finding a non-magnetic insulator that induces spin-to-charge current conversion (SCC) in graphene.

A potential candidate sorting both previous issues is $Bi_2O_3$, which has been used to obtain SCC in combination with Cu via IREE[42], as well as to modify the anomalous Hall effect in Co by inducing interfacial skew scattering[43]. In this work, using Hanle spin precession measurements, we show an unprecedented experimental demonstration of efficient SHE up to room temperature in graphene combined with $Bi_2O_3$. We observe a spin Hall angle $\theta^{gr/Bi_2O_3}$ up to 0.6% over a long spin diffusion length $\lambda_s^{gr/Bi_2O_3}$ up to 560 nm. This leads to a SCC efficiency length ($\lambda_s^{gr/Bi_2O_3} \times \theta^{gr/Bi_2O_3}$) up to 3.5 nm, larger than some of the best SCC materials such as topological insulators[44]. Our findings demonstrate a scalable graphene-based material system with efficient SCC, which is promising for building active spintronic devices.

Our device is a bilayer graphene Hall bar with $Bi_2O_3$ deposited in the central section of the cross, as shown in Figure 1a,b. Since the spin or charge current cannot be injected into or from the bulk of $Bi_2O_3$ layer since it is an insulator, bulk (I)SHE in $Bi_2O_3$ is not expected to occur in our device. Therefore, the (I)SHE and (I)REE take place only in graphene and are expected to generate (or detect) spin currents with different spin polarizations. A charge current applied along the $y$-axis of the graphene channel into the graphene/$Bi_2O_3$ region will generate the spin current propagating along the $x$-direction with out-of-plane spin polarization for the SHE[34,35] (Figure 1a) and in-plane spin polarization for the REE[34–39] (Figure 1b). This difference in the symmetry of the spin-charge interconversion allows us to distinguish the contribution from the two effects. The device used for our measurements is shown in Figure 1c. It consists of exfoliated bilayer graphene (see Note S1 for Raman characterization) shaped into three Hall bars and connected with Ti (5 nm)/Au (40 nm) contacts. In the center of the first and third Hall bar, a thin layer of $Bi_2O_3$ (5 nm) was deposited. On top of the main graphene channel, five $TiO_x$ barriers and ferromagnetic (FM) Co electrodes with different widths were fabricated, forming four different lateral spin valves (LSV). See Methods (See Note S9) for the fabrication details. In the main text, we discuss the measurements on the right side of the device (highlighted with a dotted black box in Figure 1c).

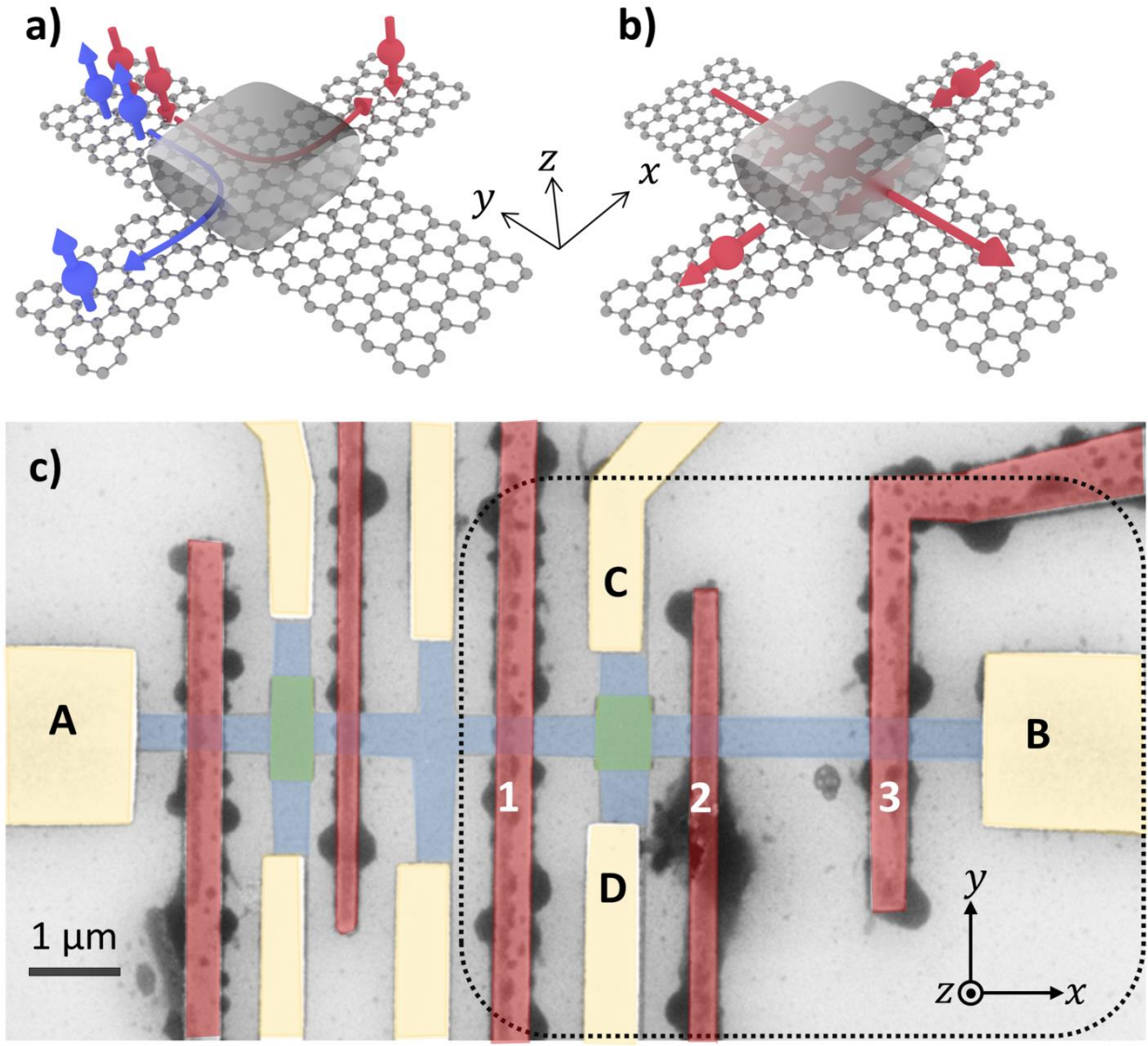

**Figure 1.** (a,b) Sketch of two possible spin-charge interconversion scenarios in graphene induced by $Bi_2O_3$ on top. A charge current applied to the graphene/$Bi_2O_3$ arm (along $y$) results in (a) a spin current with out-of-plane spin polarization (along $z$) for the SHE scenario and/or (b) a spin accumulation with in-plane spin polarization (along $x$) for the REE scenario. In both cases, a spin current diffuses in the graphene channel (along $x$). (c) False-colored scanning electron microscope image of our device. Graphene (blue) was shaped into three Hall bars with Ti/Au contacts (yellow) at the end. $Bi_2O_3$ (green) was deposited on two of the graphene Hall bars after etching the graphene flake. Co/$TiO_x$ electrodes (red) were placed on the graphene channel. The measurements explained in the main text were performed using the electrodes inside the black dotted box.

Prior to SCC measurements, we quantify the spin transport properties of our graphene. For this purpose, we studied the pristine graphene region using a reference LSV (between Co electrodes 2 and 3). An electrical current ($I = 10\ \mu A$) is applied between the FM injector (Co electrode 3) and graphene (Ti/Au contact B). This creates a spin accumulation at the Co/graphene interface, which then diffuses in the graphene channel to both sides of the Co electrode. On the left side, a pure spin current diffuses towards the FM detector, creating a nonlocal voltage ($V_{NL}$) measured between the FM (Co electrode 2) and graphene (Ti/Au contact A), which changes sign with the relative magnetization orientation of the FM electrodes (parallel, P, or antiparallel, AP). From this, the nonlocal resistance ($R_{NL} = V_{NL}/I$) is calculated. Due to differences in the shape anisotropies, the coercive fields of the FM injector and detector are different, whereas the easy axis for both electrodes lies along the $y$-axis. By applying a magnetic field along this axis ($B_y$), the two FM can be set P or AP to each other. Figure 2a shows the $R_{NL}$ vs. $B_y$ measurement (black circles) at room temperature, where a difference of 3.8 Ω between the two states ($R_{NL}^P - R_{NL}^{AP}$) is obtained. In order to quantify the spin transport properties of the pristine graphene and the spin injection and detection efficiencies of the FM electrodes, we performed standard Hanle precession measurements, as shown in the schematic diagram in Figure 2b. First, we set the magnetization of the Co electrodes P to each other along the easy axis, and then sweep a magnetic field along the in-plane hard axis ($\pm B_x$) while measuring $R_{NL}$. $B_x$ modifies the spin polarization of the spin current reaching the detector in two ways: 1) At low fields, spins in the graphene channel precess along the $y - z$ plane; 2) At larger fields, the magnetizations of the Co electrodes start to rotate towards the field direction, varying the spin polarization of the injected spins from the $y-$ to the $x-$direction. Above the saturation field, since the spin polarization of the injected spins and the magnetic field are parallel, no spin precession occurs in the graphene channel, resulting in the saturation of $R_{NL}$. We repeat the same measurement after setting the Co electrodes AP to each other. Figure 2c shows these $R_{NL}$ vs. $B_x$ symmetric Hanle curves for initial P (red) and AP (blue) configurations at room temperature. As the contribution to $R_{NL}$ due to the variation of the Co magnetization along $x$-direction is the same for both cases, subtracting the two curves removes this contribution. The average of the difference between the two curves ($\Delta R_{NL} = (R_{NL}^P - R_{NL}^{AP})/2$) corresponds to the pure spin precession signal and its variation with $B_x$ is shown in Figure 2d (black circles). By fitting this curve using the solution of the Bloch equation (see Note S7.1), the spin lifetime of the pristine graphene, $\tau_s^{gr} = 217 \pm 16$ ps, the spin diffusion constant of the pristine graphene, $D_s^{gr} = (9.03 \pm 0.26) \times 10^{-3}$ m²/s (yielding a spin diffusion length, $\lambda_s^{gr} = \sqrt{D_s^{gr} \tau_s^{gr}} = 1.4 \pm 0.1$ μm), and the spin polarization of the Co/graphene interface, $P = 6.0 \pm 0.1\%$, are obtained for room temperature. From the same measurements, the rotation of the Co magnetization with $B_x$ can also be extracted (Note S7.1.1). We found that the Co magnetization saturates along the $x-$axis at $B_x \geq 0.3$ T. We also determine the carrier density in graphene at room temperature ($n = +3.26 \times 10^{12}$ cm⁻²) from Hall measurements (Note S6).

To quantify the spin transport properties of the graphene/$Bi_2O_3$ region, we perform the same measurements using the LSV between Co electrodes 1 and 2. Figure 2a,d (green circles) show the $R_{NL}$ vs. $B_y$ and $\Delta R_{NL}$ vs. $B_x$ measurements at room temperature, respectively, which are compared to the measurements in the reference LSV (black circles). It is worth stressing that $\Delta R_{NL}$ is ~5 times smaller in the LSV with graphene/$Bi_2O_3$ region. This may indicate that the spin relaxation in the graphene/$Bi_2O_3$ region is stronger compared to the pristine graphene region. In graphene with proximity induced SOC, spin lifetime anisotropy is possible[27] causing modifications in Hanle spin precession measurement curves[28,29]. However, we did not observe any such modification (see also Note S3). We fitted the $\Delta R_{NL}$ vs. $B_x$ curve to the solution of the Bloch equation. For an accurate fitting, the spin transport channel was divided into 5 regions (details are described in Note S7.2). The spin lifetime and spin diffusion constant in the pristine graphene regions and the spin polarization of the Co/graphene interface are assumed to be the ones obtained from the measurements of the reference LSV. To simplify the analysis, the spin diffusion constant is assumed to be the same for both pristine graphene and graphene/$Bi_2O_3$ regions. With these considerations, we obtained the spin lifetime $\tau_s^{gr/Bi_2O_3} = 21 \pm 9$ ps for graphene/$Bi_2O_3$ at room temperature, which corresponds to a spin diffusion length $\lambda_s^{gr/Bi_2O_3} = 435 \pm 186$ nm.

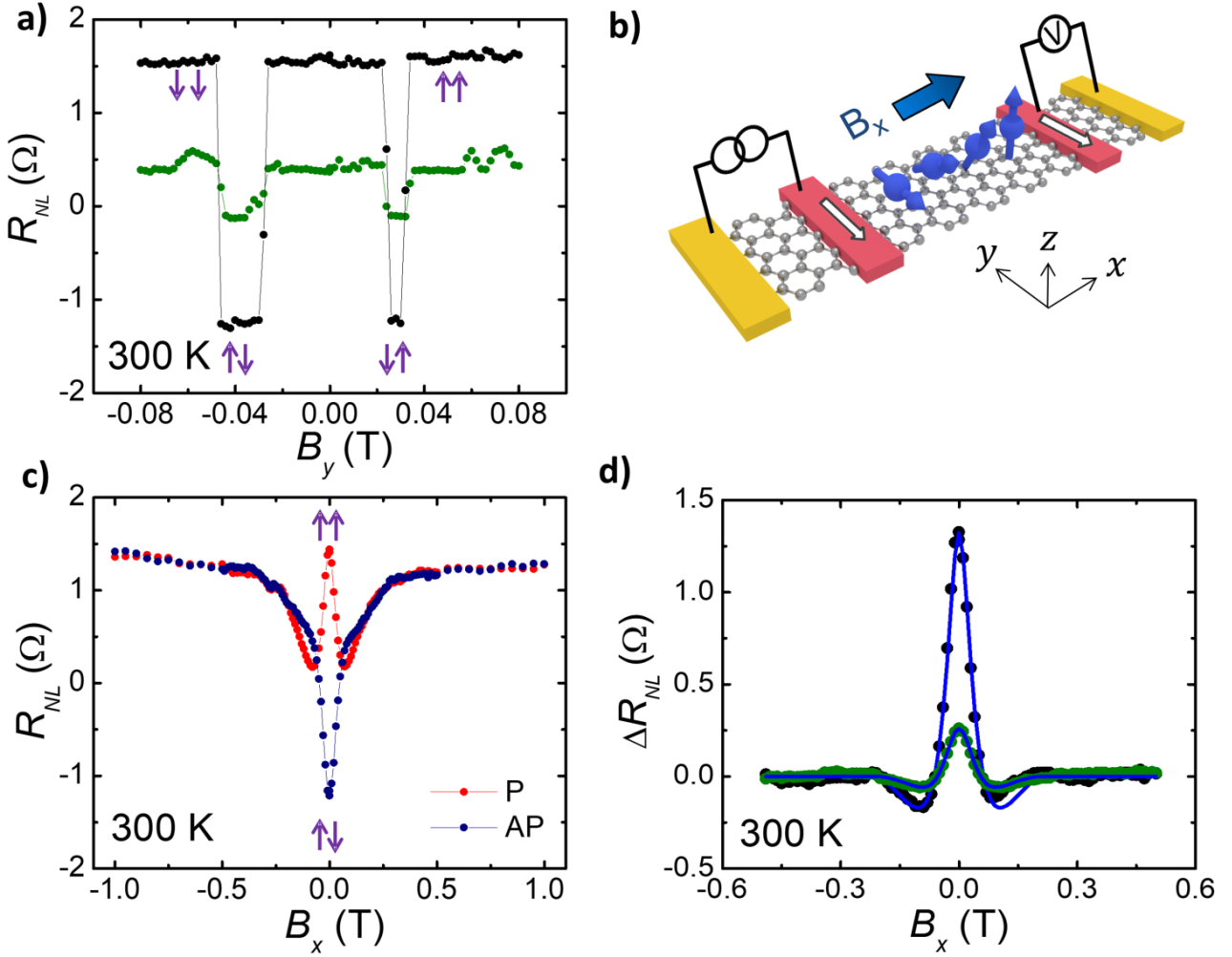

**Figure 2.** Spin transport characterization. (a) Nonlocal resistance as a function of the magnetic field applied along the easy axis of the FM electrodes ($B_y$) measured at 300 K. The black and green circles correspond to the reference LSV (between Co electrodes 2 and 3) and the LSV with $Bi_2O_3$ on top (between Co electrodes 1 and 2), respectively. The arrows represent the relative magnetizations of the two FMs. (b) Sketch of the nonlocal Hanle spin precession measurement in the reference LSV. The magnetic field is applied along the in-plane hard axis ($B_x$), causing the spin precession along the $y - z$ plane. (c) Nonlocal symmetric Hanle curves measured at 300 K at the reference LSV by applying $B_x$ for the initial parallel (red) and antiparallel (blue) configuration of the FM electrodes. (d) Net spin precession signal $\Delta R_{NL} = (R_{NL}^P - R_{NL}^{AP})/2$ calculated using the data in panel c for the reference LSV (black circles) and for the LSV with $Bi_2O_3$ on top (green circles), with the corresponding fit to the solution of the Bloch equations (blue solid line).

Once we confirmed that the spin current relaxes faster in the graphene/$Bi_2O_3$ region than in the pristine graphene, we performed the spin-to-charge conversion experiments. The electrical current ($I$= 10 µA) is applied between Co electrode 1 and Ti/Au contact A. The pure spin current that reaches the graphene/$Bi_2O_3$ region is expected to be converted into a charge current creating a nonlocal voltage, $V_{NL}$, measured across the perpendicular graphene/$Bi_2O_3$ arm (between Ti/Au contacts C and D) as shown schematically in Figure 3a. We normalize this voltage to the injected current, $R_{NL} = V_{NL}/I$. As explained in Figure 1a,b, it is possible in our device to obtain SCCs with in-plane spin polarization due to IREE or out-of-plane spin polarization due to ISHE in graphene. To identify which mechanism contributes to the SCC, we performed $R_{NL}$ vs. $B_x$ measurements. As explained above, $B_x$ causes a variation of the spin polarization along the $x$-direction via the rotation of the magnetization of the Co injector and along the $y - z$ plane via the spin precession. Therefore, the shape of $R_{NL}$ vs. $B_x$ is expected to be proportional to the variation of the Co magnetization if it arises from IREE, whereas it is expected to follow the spin precession if it comes from ISHE. For the former case, a linearly varying curve that saturates at a maxima or minima above the $\pm B_x$ saturation fields independently of the initial Co injector magnetization is expected[34–41] For the latter case, a Hanle spin precession curve with $R_{NL}$ exhibiting either a maximum or minimum at certain value of $\pm B_x$ at which the spins reaching the graphene/$Bi_2O_3$ region point out-of-plane (z) and diminishing to zero above $\pm B_x$ saturation field is expected. This plot should be antisymmetric with $B_x$ and opposite for the two opposite initial magnetization of the Co injector as the spin precession will be along opposite z-directions for the two cases[34,35]. The measurements were performed in four steps: For the first two steps, $R_{NL}$ was measured by applying $B_x$ from 0 to 1 T and 0 to -1 T by setting the initial magnetization of the Co injector along the $+y$ −axis ($R_{NL}^\uparrow$). Subsequently, the same measurements were repeated for an initial Co magnetization set along the $-y$ −axis ($R_{NL}^\downarrow$). The $R_{NL}$ vs. $B_x$ measurements at 300 K is shown in Figure 3b, where we clearly observe antisymmetric Hanle curves. Also, the two plots for opposite initial magnetizations ($R_{NL}^\uparrow$ and $R_{NL}^\downarrow$) show opposite trends at low fields. Therefore, we unambiguously confirm SCC with out-of-plane spins in our device due to ISHE in the graphene/$Bi_2O_3$ region. In contrast, we did not observe any SCC signal due to IREE in our experiment, evidencing that Rashba SOC is not dominant in our system. By interchanging the voltage and current terminals, the spin current is now injected from the graphene/$Bi_2O_3$ region and detected by the FM electrode (Figure 3d), confirming the direct SHE, which fulfills the Onsager reciprocity condition. We also performed experiments in the second graphene/$Bi_2O_3$ region on the left side of the device shown in Figure 1c and obtained similar results (Note S5), confirming the reproducibility of our measurements. This also proves the ability to fabricate multiple graphene-based spin Hall devices in the same chip under the same fabrication conditions. By averaging the difference between the $R_{NL}^\uparrow$ and $R_{NL}^\downarrow$ curves ($R_{SCC} = (R_{NL}^\uparrow - R_{NL}^\downarrow)/2$), the net antisymmetric spin precession signal can be extracted. The resulting $R_{SCC}$ vs. $B_x$ curves at 300 K is shown in Figure 3c. We fitted this curve to the solution of the Bloch equation (details of the model used for the fitting are discussed in Note S9). Here, the net SCC signal depends on $P$, $\tau_s^{gr}$, $D_s^{gr}$, and $\lambda_s^{gr/Bi_2O_3}$. By using the values of these parameters calculated from the LSV measurements (Figure 2d), the spin Hall angle of the graphene/$Bi_2O_3$ region ($\theta^{gr/Bi_2O_3}$) can be extracted. At room temperature, we obtained $\theta^{gr/Bi_2O_3} \sim 0.10 \pm 0.05\%$, corresponding to a SCC efficiency length ($\lambda_s^{gr/Bi_2O_3} \times \theta^{gr/Bi_2O_3}$) of 0.4±0.2 nm.

To understand the variation of the SHE with temperature, we performed $R_{SCC}$ vs. $B_x$ measurements at different temperatures between 10 K and 300 K. As shown in Figure 4a (see also Note S4), the amplitude of the SCC signal increases with decreasing temperature. Since the SCC conversion voltage depends not only on the spin Hall efficiency but also on the channel resistance and the spin transport properties of graphene, we need to probe the temperature dependences of the latter two to understand the actual temperature dependence of the SHE. The sheet resistances of the pristine graphene channel and the graphene channel with $Bi_2O_3$ region, calculated from the 4-point electrical measurement, increase with decreasing temperature. Interestingly, they were almost the same at each temperature (inset in Figure 4b), confirming that insulating $Bi_2O_3$ has no effect on the electrical conductivity of graphene. To obtain the spin transport properties at each temperature, the Hanle spin precession measurements across the reference LSV (Note S2) and the LSV with $Bi_2O_3$ -covered graphene (Note S3) were measured and analyzed. Using these parameters, $\theta^{gr/Bi_2O_3}$, $\lambda_s^{gr/Bi_2O_3} \times \theta^{gr/Bi_2O_3}$ and the spin Hall conductivity ($\sigma_{SH}^{gr/Bi_2O_3} = \theta^{gr/Bi_2O_3} \times R_{sq}^{gr}$) at different temperatures were calculated and are plotted in Figure 4b-d. We observe that all three parameters increase with decreasing temperature. We obtained a maximum $\theta^{gr/Bi_2O_3} = 0.6 \pm 0.1\%$ and $\lambda_s^{gr/Bi_2O_3} \times \theta^{gr/Bi_2O_3} =3.5 \pm 3.0$ nm at 10 K. Note that the large error bar in the latter value is associated to the uncertainty in the $\tau_s^{gr/Bi_2O_3}$ value at low temperatures, which is extracted with a complex five-region model that takes our device geometry into account (See Note S.7.2). The mean value of the SCC efficiency length is similar to the best SCC systems[44] such as graphene/TMD van der Waals heterostructures[35] or 2D electron gases at the surface of topological insulators[45] and LAO/STO interfaces[46].

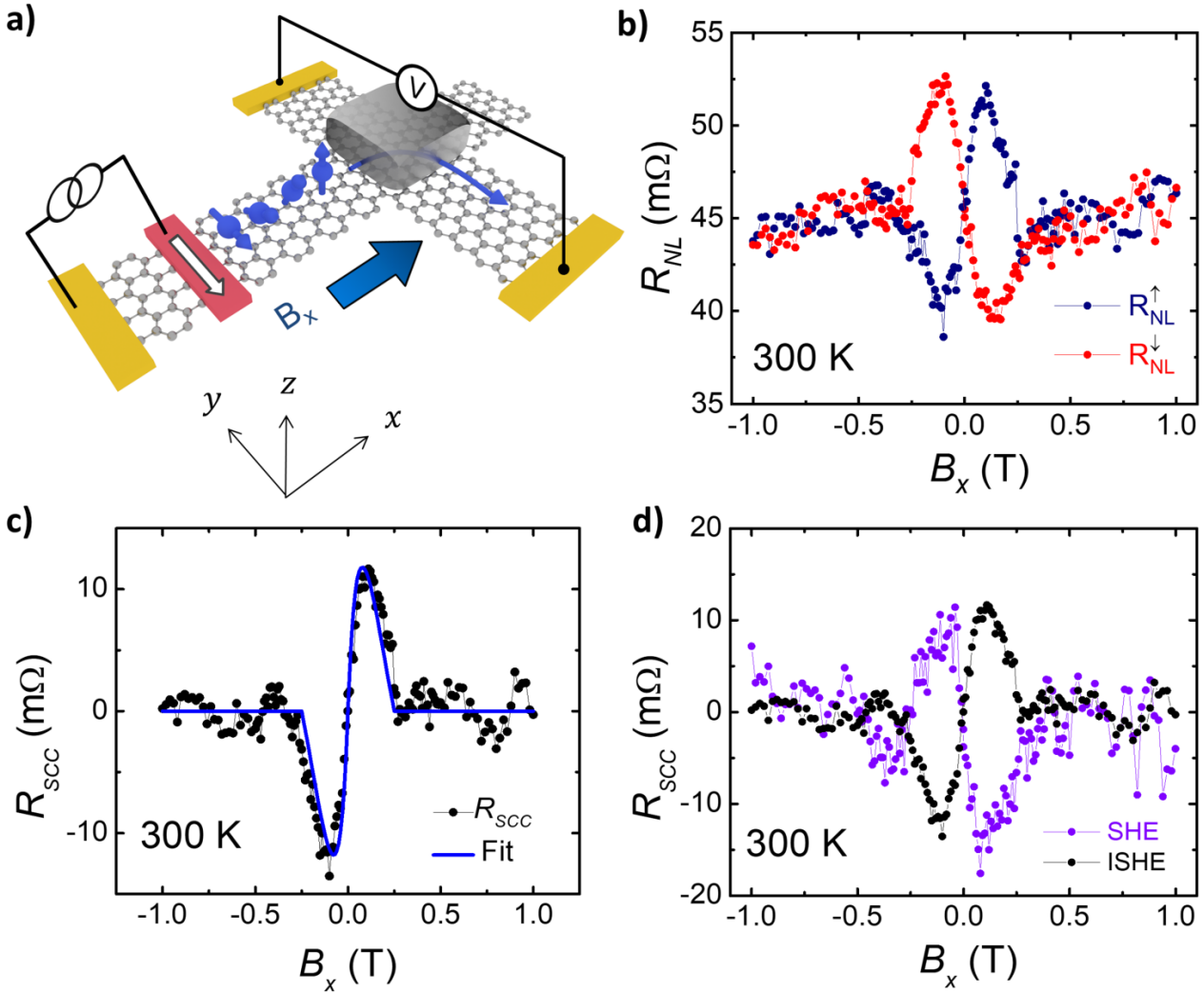

**Figure 3.** Spin-to-charge conversion measurements. (a) Sketch of the SCC measurement configuration. (b) Nonlocal spin-to-charge conversion curves at room temperature obtained by applying a charge current between Co electrode 1 and Ti/Au contact A electrode and measuring the voltage between Ti/Au contacts C and D. The magnetic field is applied along the in-plane hard axis direction ($B_x$) for the Co magnetizations initially set along positive ($R_{NL}^{\uparrow}$, blue circles) and negative ($R_{NL}^{\downarrow}$, red circles) directions along the easy axis. (c) Net antisymmetric Hanle curve (black circles) obtained by averaging the difference between the two curves ($R_{SCC} = (R_{NL}^{\uparrow} - R_{NL}^{\downarrow})/2$) from panel b, which was fitted to the solution of the Bloch equation (blue curve). (d) The net antisymmetric Hanle curve, $R_{SCC}$ vs. $B_x$, measured by swapping the current and voltage contacts at room temperature and thus corresponding to charge-to-spin conversion (violet circles). For comparison, the reciprocal spin-to-charge conversion curve in panel c is plotted again (black circles).

To date, the only unambiguous experimental observations of SHE in graphene were in proximity with TMDs[34–36]. For such heterostructure, the strong SOC and the breaking of the in-plane inversion symmetry in the honeycomb structure of the TMD leads to a large valley-Zeeman coupling, which is imprinted into graphene by proximity effect. This leads to an alteration of the spin texture of graphene states which display out-of-plane polarization, at the origin of a large SHE[19,22,32,33]. Here, we observe SHE in graphene using a completely different system. Due to the expected polycrystallinity of $Bi_2O_3$ on graphene, it is difficult to argue from the experiment whether the measured SHE is either of intrinsic or extrinsic nature since both options seem possible. The first possibility would correspond to an atomic arrangement of the $Bi_2O_3$ atoms breaking the sublattice symmetry and giving rise to a SOC term similar to the valley-Zeeman in TMD/graphene. Such symmetry breaking could also be induced by adatoms or very small impurities that may locally alter the band features of graphene. However, the most likely scenario seems to be the extrinsic one: inhomogeneities at the interface that give rise to long-range potentials induce the SHE without global alteration of the graphene band structure. Distinguishing between the intrinsic and the extrinsic mechanisms of the SHE in graphene is a challenging problem, as the changes made to the band structure are in the range of meV, which is in the range of disorder in typical graphene devices[25]. Theoretical studies predict that SHE in graphene due to both proximity effect[32] and adatoms[14,17] increases while decreasing the temperature, making it hard to experimentally distinguish between the two mechanisms using the temperature-dependent results shown in Figure 4. For this reason, the analysis used in metals[43,47,48] to quantify the contribution of the different mechanisms cannot be safely applied in graphene. The recent results of Benitez et al.[35], showing the coexistence of SHE, REE and spin lifetime anisotropy in a graphene/TMD system, appear contradictory with the theoretical predictions limited to intrinsic effects[22,32,33] and rather suggest the presence of additional extrinsic sources at the origin of SHE. While the extrinsic mechanism is the most probable scenario in our experiment, further theoretical studies are required to understand it in detail.

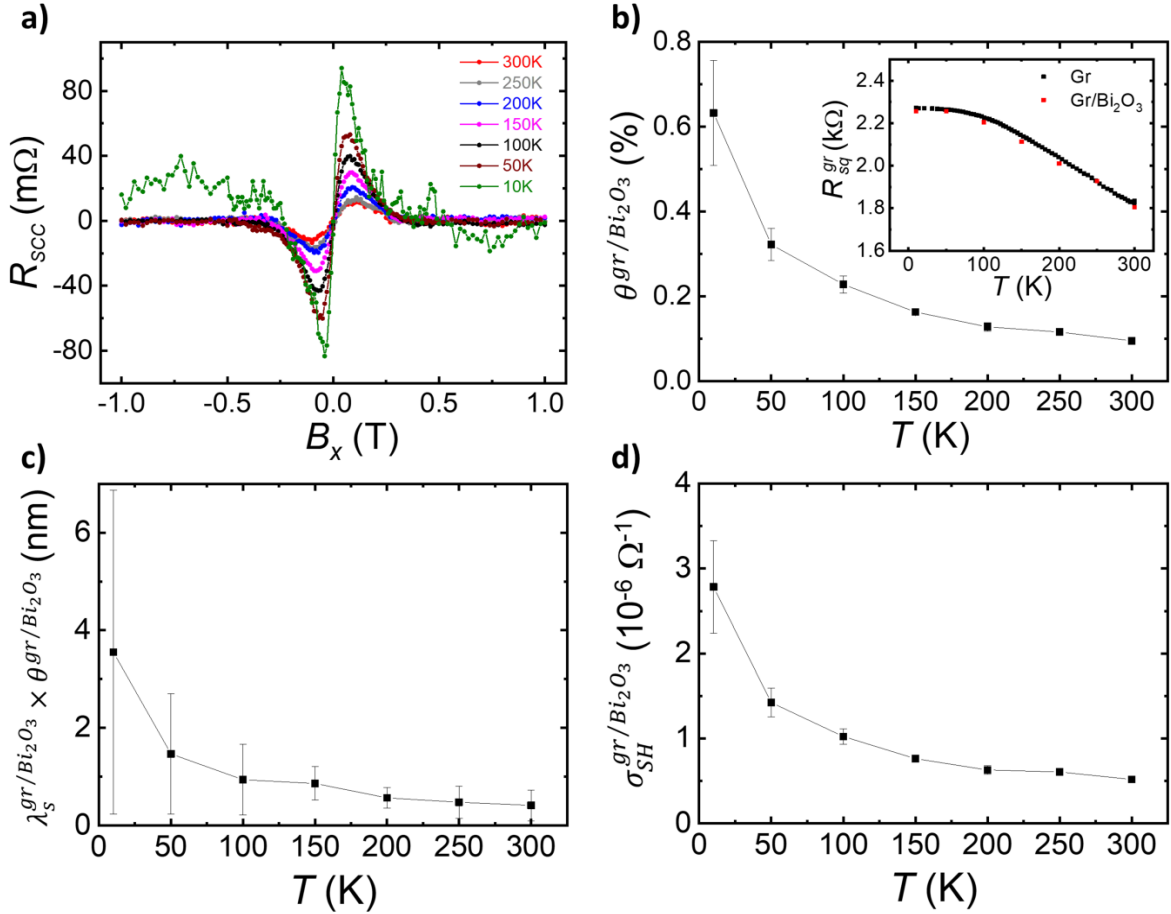

**Figure 4.** Temperature dependence of the spin-to-charge conversion. (a) Antisymmetric Hanle curves measured at different temperatures. (b) Temperature dependence of the spin Hall angle of the graphene/$Bi_2O_3$ region ($\theta^{gr/Bi_2O_3}$). Inset: Sheet resistances ($R_{sq}^{gr}$) of the pristine graphene channel (black) and the graphene channel with $Bi_2O_3$ region (red) as a function of temperature. (c) SCC efficiency length ($\theta^{gr/Bi_2O_3} \times \lambda_s^{gr/Bi_2O_3}$) as a function of temperature. (d) Spin Hall conductivity of the graphene/$Bi_2O_3$ region ($\sigma_{SH}^{gr/Bi_2O_3}$) as a function of temperature.

In conclusion, using spin precession measurements, we unambiguously demonstrate the SHE in graphene combined with insulating $Bi_2O_3$, showing large SCC efficiency. While the SHE has been experimentally reported so far only in graphene/TMD van der Waals heterostructures due to an intrinsic valley-Zeeman mechanism, we show that SHE can be also obtained by combining graphene with a polycrystalline insulator. Future studies of the dependence of SHE on the carrier density, the number of graphene layers and the growth of the SOC material on top of graphene will be steps forward to understand the origin of the SHE. Our finding opens the door to explore, both theoretically and experimentally, different spin-orbit-related effects in graphene combined with a variety of SOC materials. For device applications, a graphene/oxide heterostructure holds multiple advantages. Unlike the fabrication of van der Waals heterostructures, $Bi_2O_3$ can be deposited on top of graphene using scalable deposition techniques. This would allow large-scale fabrication of devices on a single chip. As an insulator, $Bi_2O_3$ can be easily integrated to electronic circuits without short-circuiting the charge transport. Moreover, the SCC is localized in graphene since it cannot occur in the bulk of $Bi_2O_3$. All these features will be more suitable for future spintronic device applications such as graphene-based spin-orbit torque memories or spin-logic devices, where large charge or spin current densities in graphene, respectively, will be required.

- **ASSOCIATED CONTENT**

**Supporting Information**

Additional details on methods: sample fabrication and characterization, spin transport in the reference LSV and LSV with graphene/$Bi_2O_3$ region at different temperatures, spin-to-charge conversion measurements at different temperatures, reproducibility, Hall measurements to determine the carrier density of graphene, determination of the spin lifetime in the different graphene regions, determination of the spin Hall angle in the graphene/$Bi_2O_3$ region.

- **AUTHOR INFORMATION**

**Corresponding Authors**
*E-mail: f.casanova@nanogune.eu

**Notes**
The authors declare no competing financial interest.

- **ACKNOWLEDGMENTS**

The authors thank Dr. José H. Garcia and Prof. Stephan Roche for fruitful discussions and Roger Llopis for drawing the sketches used in the figures. This work is supported by the Spanish MINECO under the Maria de Maeztu Units of Excellence Programme (MDM-2016-0618) and under Project RTI2018-094861-B-100, and by the European Union H2020 under the Marie Curie Actions (794982-2DSTOP and 766025-QuESTech). N.O. thanks the Spanish MINECO for a Ph.D. fellowship (Grant No. BES-2017-07963). J.I.-A. acknowledges postdoctoral fellowship support "Juan de la Cierva - Formación" program by the Spanish MINECO (Grant No. FJC2018-038688-I).

# Supporting information

## Table of contents:

# S1. Methods: Sample fabrication and characterization

## S1.1 Device fabrication and physical characterization

First, we exfoliated graphene from bulk graphite crystals (supplied by NGS Naturgraphit GmbH) using a Nitto tape (Nitto SPV 224P) onto Si substrates with 300 nm $SiO_2$. Graphene flakes are identified by optical contrast under an optical microscope and characterized by Raman spectroscopy (Figure S1a). Graphene Hall bars were fabricated using electron-beam lithography and reactive ion etching, followed by annealing of the device at 400°C for 1 hour in ultra-high vacuum ($10^{-9}$ torr). Graphene is then connected with Ti (5 nm)/Au (40 nm) contacts fabricated using electron-beam lithography followed by thermal deposition in ultrahigh vacuum and lift-off. Using a similar procedure, a $Bi_2O_3$ thin film (5 nm) is deposited. The oxide layer was deposited from stoichiometric $Bi_2O_3$, although the deposited layer might be different, i.e., $BiO_x$. Using electrical measurements on a thin film of $Bi_2O_3$ on another substrate fabricated during the same deposition, we confirmed that $Bi_2O_3$ is insulating. Finally, $TiO_x$ tunnel barriers, grown by depositing 3 Å of Ti followed by oxidation in air, and Co electrodes (35 nm) were fabricated on top of the graphene channel. The widths of the Co electrodes vary between 250 nm to 400 nm. The interface resistances of 2.5 kΩ, 22 kΩ, and 3.6 kΩ were obtained for Co electrodes 1, 2 and 3 respectively, all of them larger than the sheet resistance of graphene at different temperatures from 300 K to 10 K. From atomic force microscopy images (Figure S1b), we confirmed the actual thickness of the $Bi_2O_3$ layer to be ~ 6.2 nm, much larger than its roughness ~ 1.5 nm, ensuring the full coverage of graphene by $Bi_2O_3$ (see Note S1). After the measurements, the exact dimensions of the devices were extracted by scanning electron microscopy images.

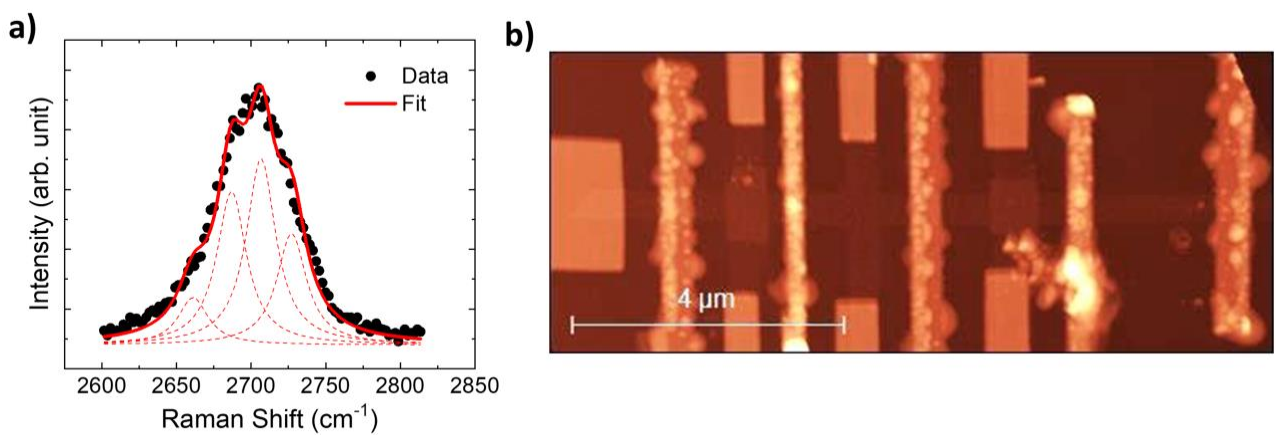

**Figure S1.** a) Raman spectroscopy of the graphene flake prior to the etching process to determine the thickness of the flake. By fitting the 2D peak into four Lorentzians[1], each with a full width half maximum (FWHM) of ~24 $cm^{-1}$, we can determine the flake to be bilayer graphene. b) The atomic force microscopy image of our sample. From this, the actual thickness of $Bi_2O_3$ at the center of the graphene Hall bar was measured to be ~ 6.2 nm, much larger than its r.m.s. roughness (~ 1.5 nm), which confirms the graphene is fully covered by $Bi_2O_3$.

## S1.2. Electrical measurements

The transport measurements were performed in a Physical Property Measurement System by Quantum Design, with a Keithley 2182 nanovoltmeter and a 6221 current source in a dc reversal mode at temperatures ranging from 10 K to 300 K. We applied out-of-plane and in-plane magnetic fields with a superconducting solenoid magnet and a rotatable sample stage.

## S2. Spin transport in the reference LSV at different temperatures

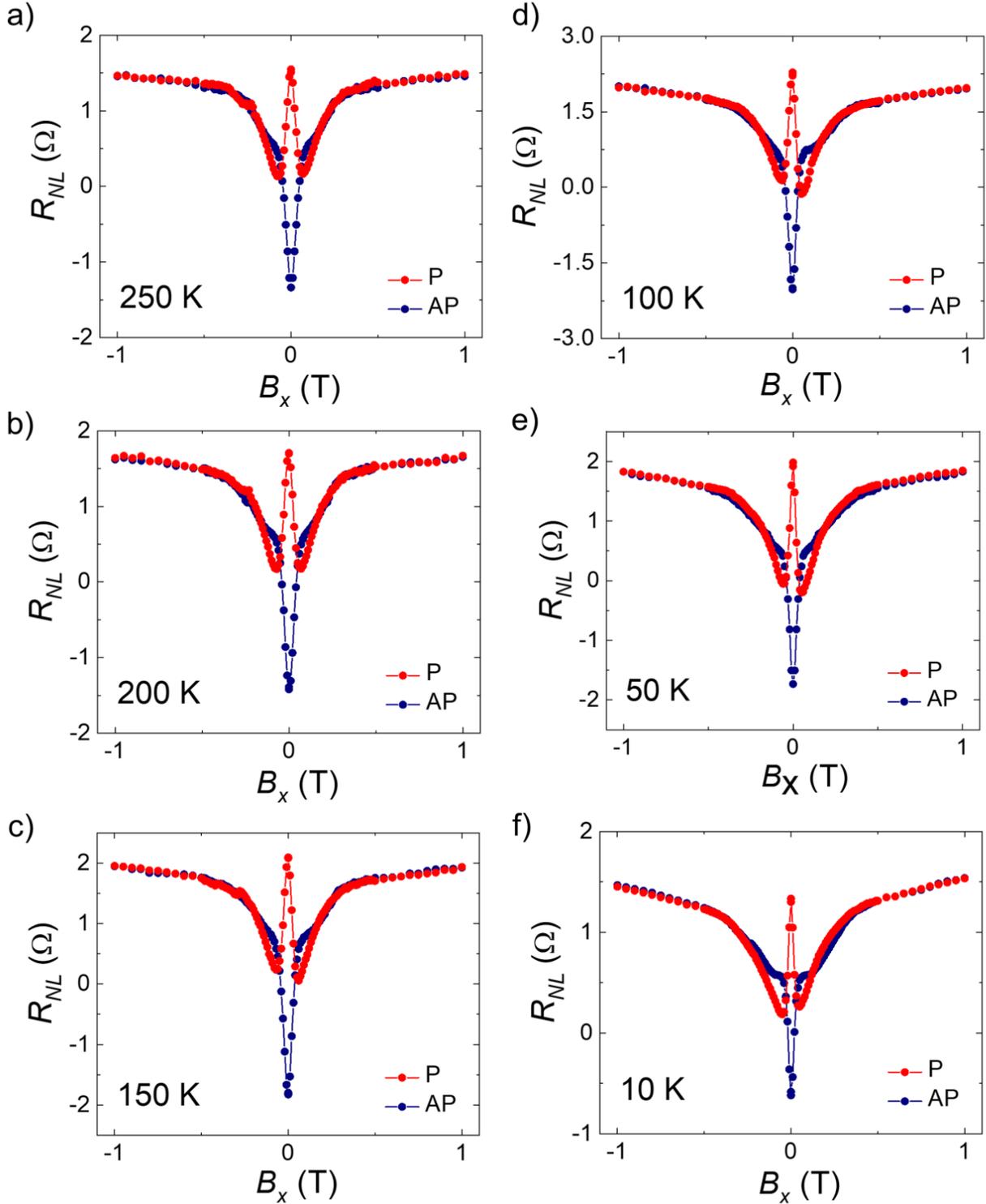

**Figure S2.** Nonlocal Hanle spin precession measurements in the reference LSV, using $V_{2,A}I_{3,B}$ terminal configuration shown in Figure 1c of the main text (also in Figure S5a), performed by applying $B_x$ for initial parallel (P, red curve) and antiparallel (AP, blue curve) states of the Co electrodes, at temperature (a) 250 K, (b) 200 K, (c) 150 K, (d) 100 K, (e) 50 K, and (f) 10 K.

## S3. Spin transport in the LSV with graphene/Bi$_2$O$_3$ region at different temperatures

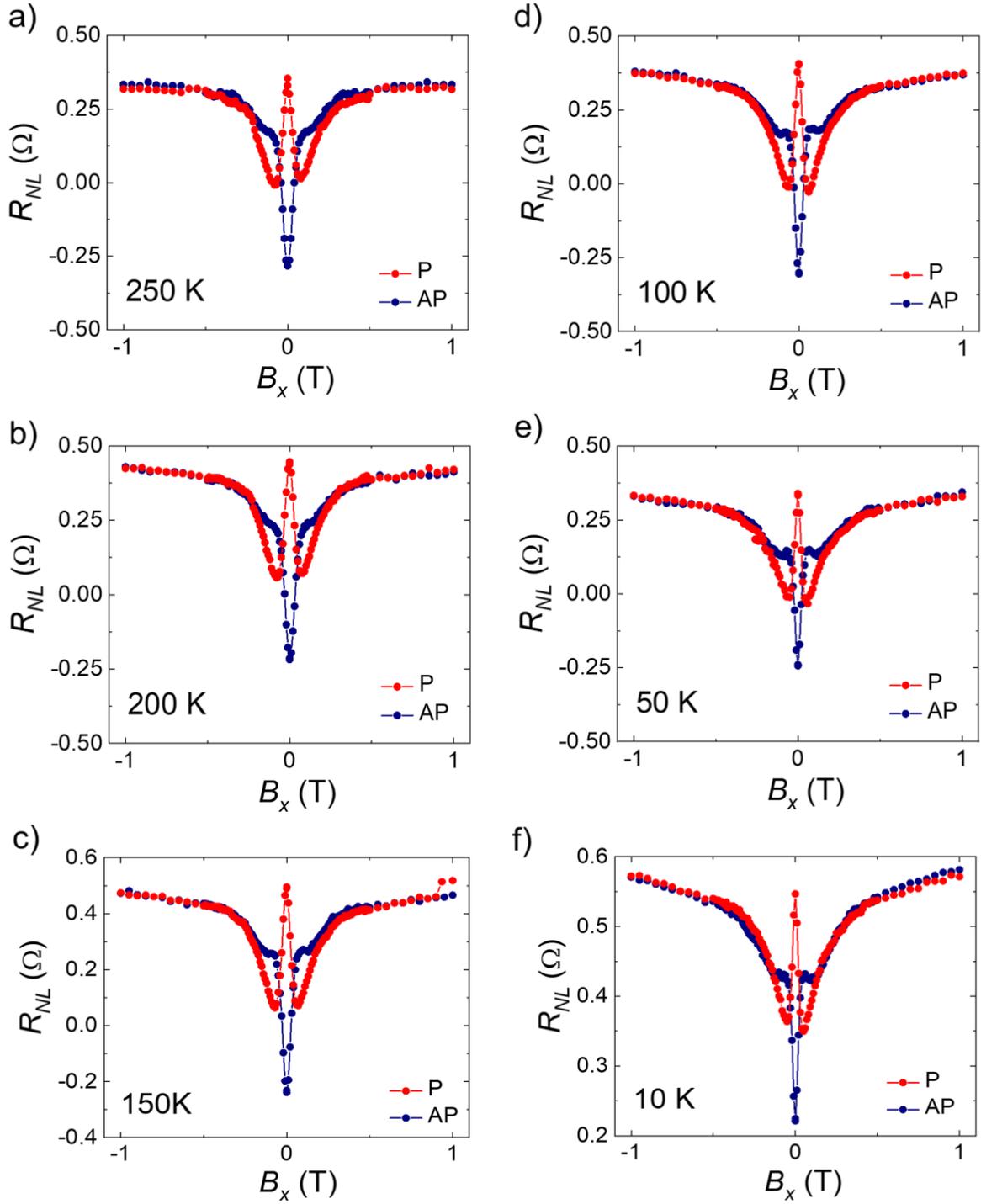

**Figure S3.** Nonlocal Hanle spin precession measurements in the LSV with Bi$_2$O$_3$ on top, using $V_{2,B}I_{1,A}$ terminal configuration shown in Figure 1c of the main text (also, in Figure S5a), performed by applying $B_x$ for initial parallel (P, red curve) and antiparallel (AP, blue curve) states of the Co electrodes, at temperature (a) 250 K, (b) 200 K, (c) 150 K, (d) 100 K, (e) 50 K, and (f) 10 K.

## S4. Spin-to-charge conversion measurements at different temperatures

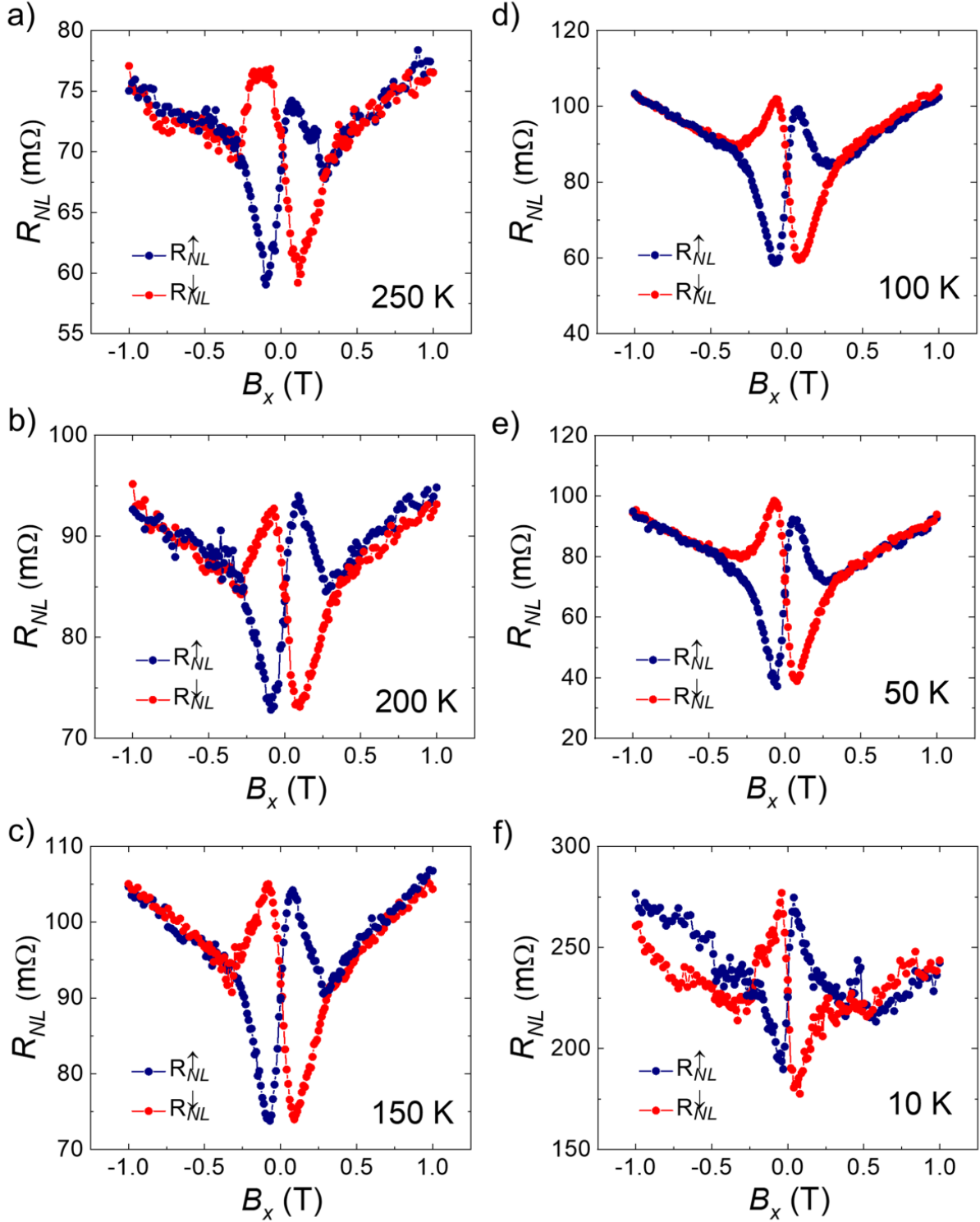

**Figure S4.** Nonlocal spin-to-charge conversion signal ($R_{NL}$), using $V_{C,D}I_{1,A}$ terminal configuration shown in Figure 1c of the main text (also in Figure S4a) as a function of the magnetic field applied along the in-plane hard axis direction ($B_x$), for initial magnetization of the Co electrode saturated along positive ($R_{NL}^{\uparrow}$, blue circles) and negative ($R_{NL}^{\downarrow}$, red circles) easy axis ($y$-direction) at temperature (a) 250 K, (b) 200 K, (c) 150 K, (d) 100 K, (e) 50 K, and (f) 10 K. A parabolic background signal was also observed in most of the measurements most likely due to the magnetoresistance in graphene.

## S5. Reproducibility

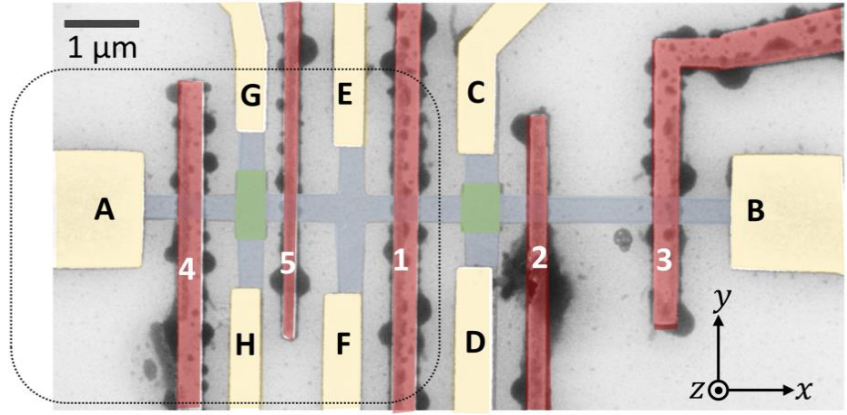

**Figure S5.** SEM image of the measured device also shown in Figure 1c of the main text with different contacts used for reproducibility experiments (performed using the electrodes inside the black dotted box).

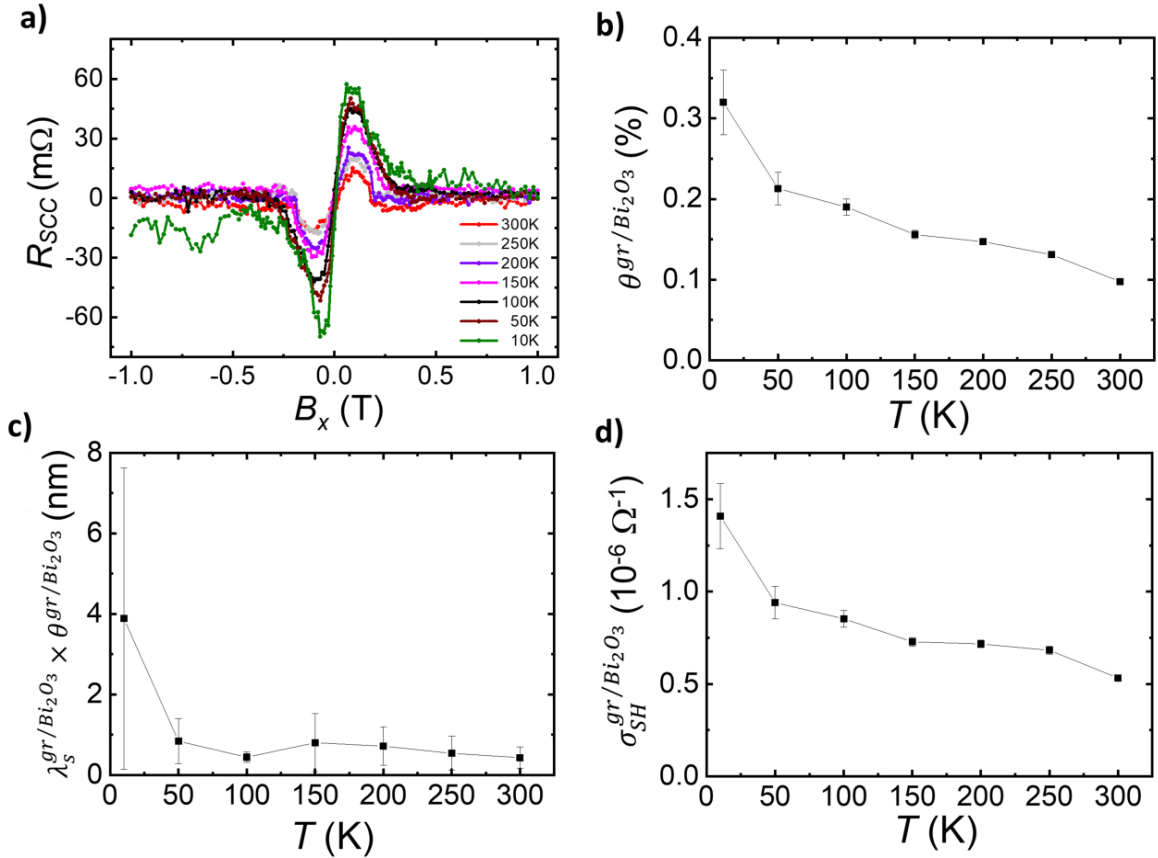

**Figure S6.** (a) Net antisymmetric Hanle curve ($R_{SCC}$) measured using $V_{G,H}I_{4,A}$ terminal configuration shown in figure S5. The signal decreases with increasing temperature, similar to the results obtained in the first device (Figure 4a in the main text), confirming the reproducibility of our results. (b) Temperature dependence of the spin Hall angle of the graphene/Bi$_2$O$_3$ region ($\theta^{gr/Bi_2O_3}$) (c) SCC efficiency length ($\theta^{gr/Bi_2O_3} \times \lambda_s^{gr/Bi_2O_3}$) as a function of temperature. (d) Spin Hall conductivity of the graphene/Bi$_2$O$_3$ region ($\sigma_{SH}^{gr/Bi_2O_3}$) as a function of temperature.

## S6. Hall measurements to determine the carrier density of graphene

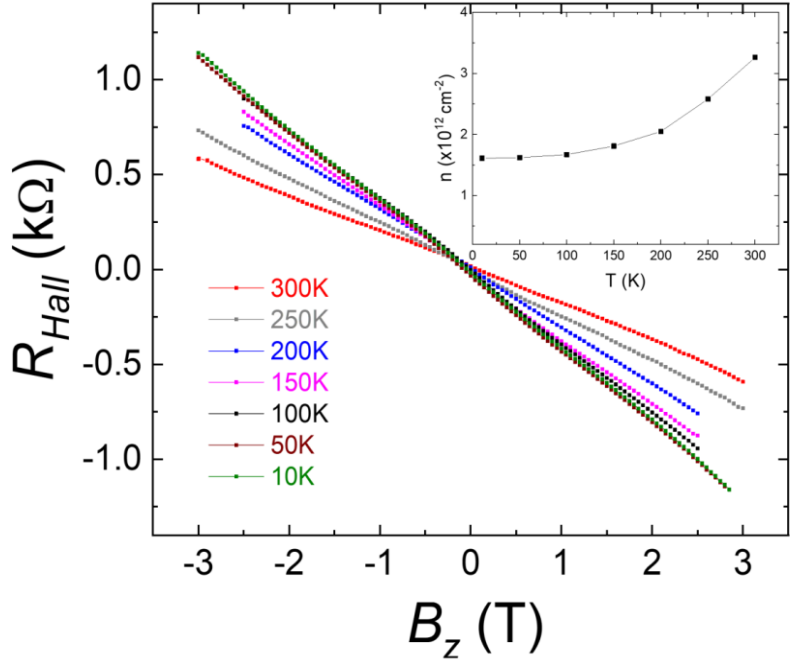

**Figure S7.** Hall resistance measured at different temperatures (using $V_{E,F}I_{A,B}$ terminal configuration with out-of-plane magnetic field). The slope of the plot gives the Hall coefficient $= -\frac{1}{ne}$, where $n$ is the carrier density. Inset: The variation of carrier density as a function of the temperature calculated from the Hall measurements.

## S7. Determination of the spin lifetime in the different graphene regions

### S7.1. Pristine graphene region

To analyze quantitatively the results described in this work, we have assumed that our device has two different and homogeneous regions: The pristine graphene and the $Bi_2O_3$-covered graphene. To determine the spin lifetime $\tau_s^{gr}$ and diffusion coefficient $D_s^{gr}$ of the pristine region, as well as the contact polarization of the Co electrodes $P$, we have performed Hanle spin precession measurements in this part of the device by applying a magnetic field along $x$ ($B_x$) while gauging the nonlocal resistance $R_{NL}$. The application of $B_x$, which is perpendicular to the easy axis of our Co electrodes has two effects: At low fields, it induces precession of the injected spins and, as the field increases, it also leads to the pulling of the contact magnetization towards the $x$-direction.

The spin precession component $R_S(B_x)$ has been obtained from the Bloch equation and accounts for the spin absorption at the contacts as described in Ref 2-4  Below this point we refer to $R_S(B_x)$ as the "*one-region model*".

#### S7.1.1. Contact pulling: The Stoner-Wohlfarth model
For the contact pulling, we assume that the magnetizations follow the Stoner-Wohlfarth model[5] which in our case, because $B_x$ is perpendicular to the easy axis, implies that the $x$-component of the electrode magnetization $M_x$ is proportional to $B_x/B_x^{sat}$, where $B_x^{sat}$ is the field at which the magnetization saturates along $x$. As a consequence, we can determine the magnetization angle $\theta_M$ with respect to the easy axis $y$: $\theta_M = \arcsin(B_x/B_x^{sat})$ for $|B_x| < B_x^{sat}$ and $\theta_M = \pm\pi/2$ for $|B_x| > B_x^{sat}$ where the $\pm$ distinguishes positive and negative $B_x$, respectively.

Because the electrodes are defined with different widths, different contacts have different coercivities. Consequently, the magnetizations can be addressed individually, and it is possible to prepare them in both parallel and antiparallel configurations. This gives rise to different nonlocal spin signals $R_{NL}^{P(AP)}$ for the parallel (antiparallel) configuration.

Considering the contact pulling, the nonlocal spin signal can be written as:

$$R_{NL}^{P(AP)} = \pm R_S(B_x)\cos^2(\theta_M) + R_\parallel \sin^2(\theta_M)$$

where $R_\parallel$ is the spin signal at $B_x = 0$. Subtraction of both curves leads to:

$$\Delta R_{NL} = \frac{R_{NL}^P - R_{NL}^{AP}}{2} = R_S(B_x)\cos^2(\theta_M) \quad (S1)$$

which is commonly fit to determine the spin transport properties of graphene[4].
Addition of both curves leads to:

$$R_{avg} = \frac{R_{NL}^P + R_{NL}^{AP}}{2} = R_\parallel \sin^2(\theta_M)$$

from here, since $R_\parallel$ is a constant, we can normalize $R_{avg}$ to obtain $\theta_M$ as a function of the magnetic field.

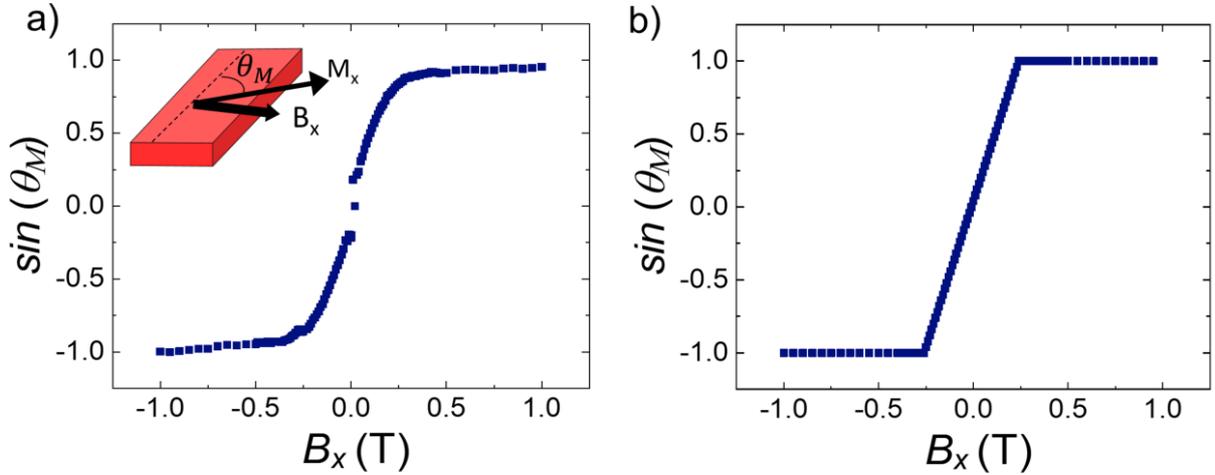

**Figure S8.** (a) sin $(\theta_M)$, where $\theta_M$ is the angle between Co magnetization rotated along $x$ with respect to the easy axis ($y$), as a function of $B_x$ obtained from the Hanle measurements across the reference LSV at 300 K (Figure 2c in main text) by using Equation S2. Inset: Sketch that defines the angle $\theta_M$. (b) sin $(\theta_M)$ as a function of $B_x$ using the Stoner-Wohlfarth model.

In Figure S8a we plot

$$sin(\theta_M) = sign(B_x)\sqrt{\frac{R_{avg} - R_{avg}^{min}}{R_{avg}^{max} - R_{avg}^{min}}} \quad (S2)$$

as a function of $B_x$ where $R_{avg}^{max}$ and $R_{avg}^{min}$ are the maximum and minimum values of $R_{avg}$ and $sign(B_x)$ is the sign of the applied magnetic field. Because for some reference Hanle precession curves the signal keeps increasing for fields higher than $B_{sat}$, for our analysis we have used the contact pulling from the Stoner-Wohlfarth model (Figure S8b).

**S7.1.2. Fitting of the spin precession data**

The spin precession signal ($\Delta R_{NL}$) obtained from the Hanle measurements in the reference LSV at 300 K is shown in Figure S9a together with the fit to Equation S1. The parameters $\tau_s^{gr}$, $D_s^{gr}$, and $P$ extracted from this analysis at different temperatures are plotted in Figure S9b,c.

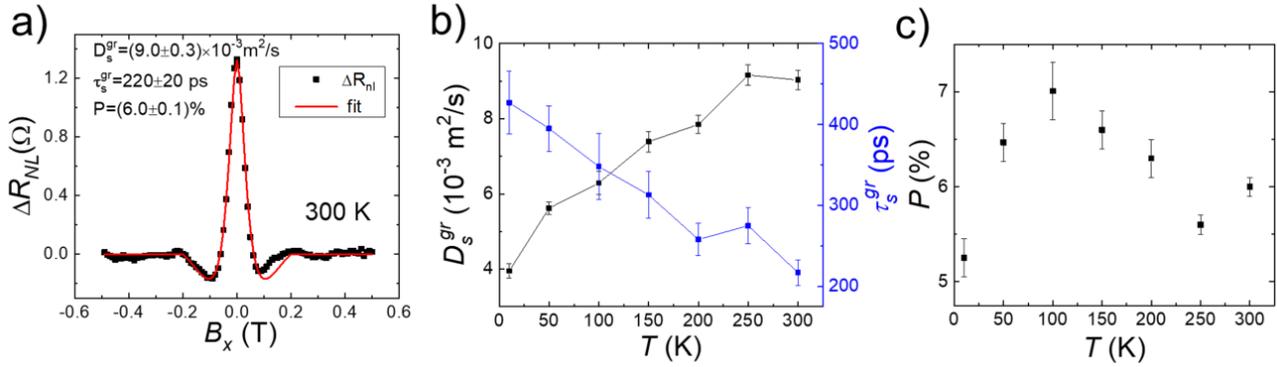

**Figure S9.** Spin transport in the pristine graphene region. (a) $\Delta R_{NL}$ at 300 K together with its fit to the solution of the Bloch equations and the extracted parameters. (b) Spin lifetime, spin diffusion constant, and (c) spin polarization obtained from the analysis shown in panel a at different temperatures.

**S.7.2. Bi$_2$O$_3$-covered region**

**S.7.2.1. One-region model**

In order to determine the spin lifetime of the Bi$_2$O$_3$-covered region, we have followed the approach introduced in Ref 6 The first step is to fit $\Delta R_{NL}$ obtained across the Bi$_2$O$_3$-covered region to obtain the effective spin transport parameters of this part of the sample. The Hanle precession curve obtained in the LSV with the Bi$_2$O$_3$-covered region at 300 K is shown in Figure S10a together with the fit to Equation S1. The effective parameters $\tau_s^{eff}$, $D_s^{eff}$, and $P^{eff}$ extracted from this analysis at different temperatures are plotted in Figure S10b,c.

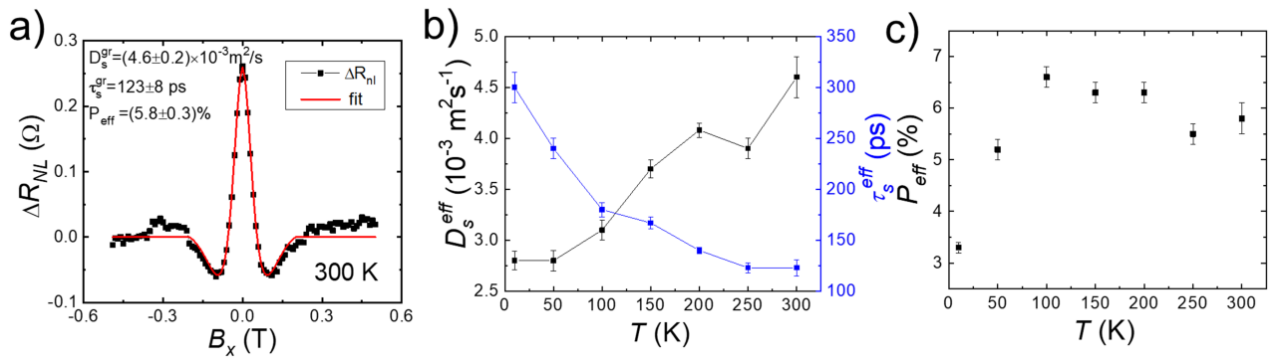

**Figure S10.** Spin transport in the Bi$_2$O$_3$-covered graphene region. (a) $\Delta R_{NL}$ at 300 K together with its fit to the solution of the Bloch equations using the one-region model and the extracted effective parameters. (b) Effective spin lifetime, spin diffusion constant, and (c) spin polarization obtained from the analysis shown in panel a at different temperatures.

### S.7.2.2. Five-region model for spin precession

The next step is to simulate the actual device geometry using the Bloch equations in the appropriate geometry and determine the spin lifetime of region $III$ ($\tau_s^{gr/Bi_2O_3}$). In our case, the geometry is shown in Fig. S10a and includes 5 different regions.

The analytical expression for spin precession in this geometry that takes into account the spin absorption by the contacts has been obtained following Refs. [3,4,7] Firstly, the solution of the Bloch equations for the spin accumulation $\mu_{y(z)}$ in the $y(z)$ direction is written as:

$$\mu_y = A\exp\left(xa_p^{gr(gr/Bi_2O_3)}\right) + B\exp\left(xa_m^{gr(gr/Bi_2O_3)}\right) + C\exp\left(-xa_p^{gr(gr/Bi_2O_3)}\right)$$
$$+ D\exp\left(xa_m^{gr(gr/Bi_2O_3)}\right)$$

$$\mu_z = iA\exp\left(xa_p^{gr(gr/Bi_2O_3)}\right) - iB\exp\left(xa_m^{gr(gr/Bi_2O_3)}\right) + iC\exp\left(-xa_p^{gr(gr/Bi_2O_3)}\right)$$
$$- iD\exp(xa_m^{gr(gr/Bi_2O_3)})$$

where $a_{p(m)}^{gr(gr/Bi_2O_3)} = \frac{\sqrt{1+(-)i\omega\tau_s^{gr(gr/Bi_2O_3)}}}{\lambda_s^{gr(gr/Bi_2O_3)}}$ and $\lambda_s^{gr(gr/Bi_2O_3)} = \sqrt{D_s^{gr(gr/Bi_2O_3)}\tau_s^{gr(gr/Bi_2O_3)}}$ with $\omega = \frac{g\mu_B B_x}{\hbar}$ the Larmor frequency, $g = 2$ the Landé factor, $\mu_B$ the Bohr magneton, and $\hbar$ the reduced Plank constant. $A - D$ are parameters to be determined from the device geometry.

The relevant boundary conditions used to determine the parameters in the five different regions are:

1. The spin accumulations $\mu_y$ and $\mu_z$ are continuous through the entire sample.
2. The spin currents, defined as $I_s^{y(z)} = \frac{W}{eR_{sq}}\frac{d\mu_{y(z)}}{dx}$, are continuous everywhere apart from $x = 0$ and $x_{det}$ where the spin injector and detector are placed.
3. At $x = 0$, the spin current has a discontinuity $\Delta I_s^y = P_i I - \frac{\mu_y(0)}{eR_{c1}}$ and $\Delta I_s^z = -\frac{\mu_z(0)}{eR_{c1}}$ where $R_{c1}$, $P_i$ and $I$ are the contact resistance, spin polarization and charge current applied through the spin injector.
4. At $x = x_{det}$, $\Delta I_s^y = -\frac{\mu_y(x_{det})}{eR_{c2}}$ and $\Delta I_s^z = -\frac{\mu_z(x_{det})}{eR_{c2}}$, where $R_{c2}$ is the contact resistance of the spin detector
5. Finally, we take into account that $\mu_{y(z)}(x \to \infty) \to 0$, which, for region $I$ implies that $C = D = 0$ and, for region $V$, $A = B = 0$.

From these boundary conditions we obtain the spin accumulation at $x = x_{det}$ as a function of $B_x$. We convert this spin accumulation in a nonlocal voltage: $V_S = P_d \mu_y(x_{det})/e$ where $P_d$ is the detector spin polarization. $V_S$ is converted in a nonlocal resistance using

$$R_S = V_S/I \qquad (S3)$$

This way we obtain the nonlocal resistance detected across the Bi$_2$O$_3$-covered region.

### S.7.2.3. Determination of $\tau_s^{gr/Bi_2O_3}$ using the five-region model

To determine $\tau_s^{gr/Bi_2O_3}$ we use the five-region model. As described above, this model depends on a long list of parameters: $D_s^{gr(gr/Bi_2O_3)}$, $\tau_s^{gr(gr/Bi_2O_3)}$, $R_{sq}^{gr(gr/Bi_2O_3)}$, and $P_{i(d)}$. The spin transport parameters of the pristine graphene region ($D_s^{gr}$, $\tau_s^{gr}$) are shown in Figure S10b. The spin polarization of the injector and detector is assumed to be identical ($P = P_i = P_d$) and it is shown in Figure S10c. The sheet resistance of pristine graphene and Bi$_2$O$_3$-covered graphene, assumed to be identical ($R_{sq}^{gr} = R_{sq}^{gr/Bi_2O_3}$), is shown in the inset of Fig. 4a in the main text. We also assume that $D_s^{gr} = D_s^{gr/Bi_2O_3}$. The relevant output parameter from our model is $\tau_s^{gr/Bi_2O_3}$, which is assumed to be isotropic because we did not observe any anisotropic feature in the spin precession data (see Figure S3). Note that here we do not determine $D_s^{gr/Bi_2O_3}$ accurately. This is because this parameter determines the shape of the shoulders of the Hanle precession curve and, in our case, these are shaped by the pulling of the contact magnetizations preventing a more accurate analysis. In contrast, $\tau_s^{gr/Bi_2O_3}$ determines the width of the central peak and can be obtained in a more reliable manner.

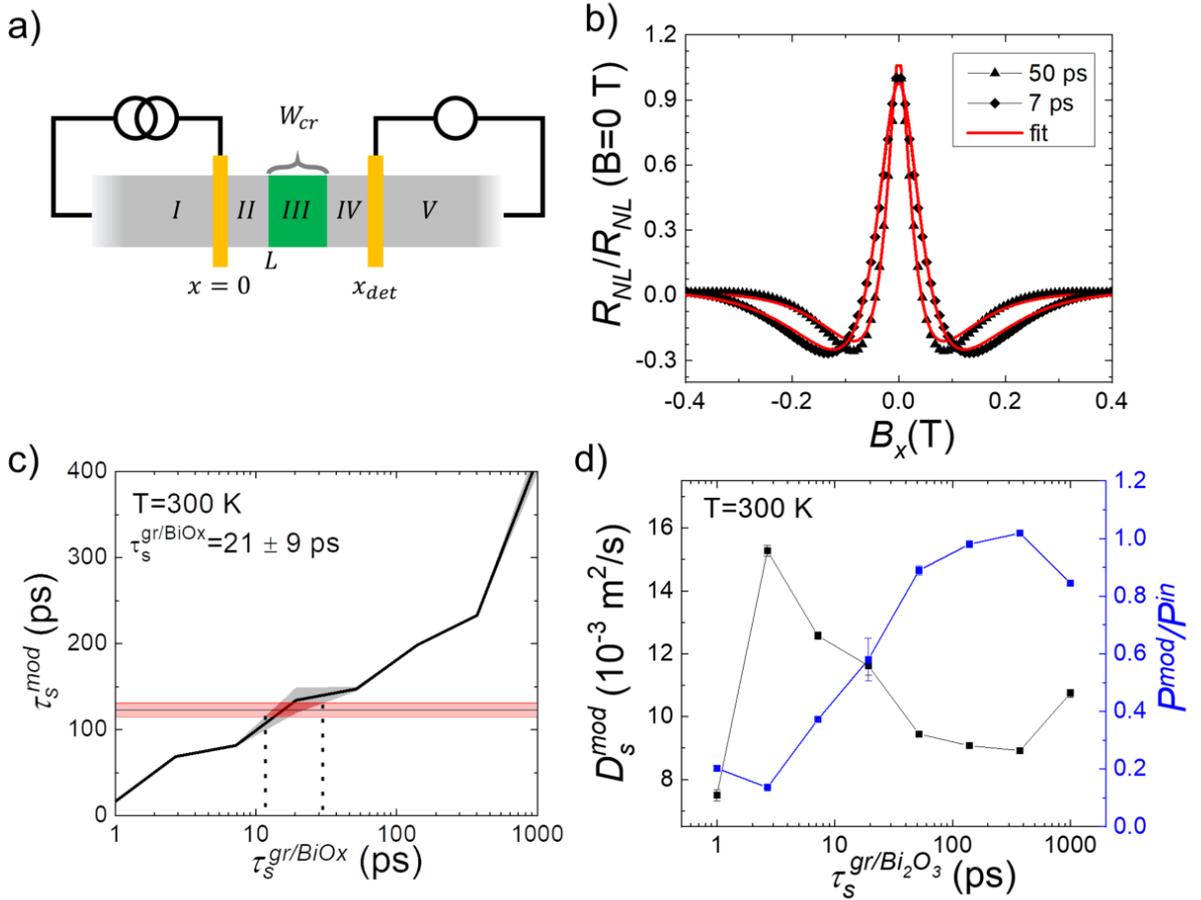

**Figure S11**. (a) Model geometry for the spin precession across the Bi$_2$O$_3$-covered region and spin-to-charge conversion. The dark yellow contacts are spin polarized, regions $I$, $II$, $IV$, and $V$ are pristine graphene, region $III$ is covered by Bi$_2$O$_3$. Regions $I$ and $V$ are semi-infinite. (b) Simulated Hanle spin precession data using the 5-region model for two cases with $\tau_s^{gr/Bi_2O_3} = 7$ and $50$ ps. The red lines correspond to the fits to the one-region model. (c) $\tau_s^{mod}$ as a function of $\tau_s^{gr/Bi_2O_3}$ (black line). The grey area is the error associated with the one-region model fit. The horizontal red line corresponds to the value of $\tau_s^{eff}$ obtained from the homogeneous fit (see Figure S10a) and has its corresponding error range. (d) $\tau_s^{mod}$ and $P^{mod}/P^{in}$, where $P^{in} = \sqrt{P_i P_d}$, that are the input spin polarizations, as a function of $\tau_s^{gr/Bi_2O_3}$.

The analysis is realized as follows: first, the Hanle precession data is simulated using Eq S3 for different values of $\tau_s^{gr/Bi_2O_3}$ (see Figure 11b), then the extracted curve is fit to the one-region model to obtain the effective parameters $\tau_s^{mod}$, $D_s^{mod}$ and $P^{mod}$. The value of $\tau_s^{gr/Bi_2O_3}$ is obtained by plotting $\tau_s^{mod}$ vs $\tau_s^{gr/Bi_2O_3}$ and finding the value of $\tau_s^{gr/Bi_2O_3}$ that gives $\tau_s^{mod} = \tau_s^{eff}$. This process is illustrated in Figure S11c for the 300 K data. We include the one-region model to determine the uncertainty range graphically taking into account the fit error of $\tau_s^{gr}$.

In Figure S11c one can see how $\tau_s^{mod}$ increases with $\tau_s^{gr/Bi_2O_3}$. This is expected as the spin lifetimes of the different regions contribute to the average lifetime of the system. In Figure S11d, a less straightforward behavior is observed for $D_s^{mod}$. In particular, for $\tau_s^{gr/Bi_2O_3} < \tau_s^{gr}$, $D_s^{mod} > D_s^{gr/Bi_2O_3}$. This is caused by the fact that the reduced $\tau_s^{gr/BiO_x}$ gives rise to a decrease of the average transport times at the Bi$_2$O$_3$-covered region, leading to less spin precession for that region. Consequently, this region behaves like a spacer and mimics the effect of an increased diffusivity that would also get the spins faster to the detector. The result from $\tau_s^{gr/Bi_2O_3} = 1$ ps is caused by a bad fit at high field that underestimates $D_s^{mod}$. Finally, the spin polarizations decrease as $\tau_s^{gr/Bi_2O_3}$ decreases. This is expected from the fact that the Bi$_2$O$_3$-covered region behaves as a spin sink when $\tau_s^{gr/Bi_2O_3} < \tau_s^{gr}$. This leads to reduced signals that result in a decrease of $P^{mod}/P^{in}$.

The above mentioned analysis has been performed at all the measured temperatures, leading to the extraction of $\tau_s^{gr/Bi_2O_3}$ with its corresponding uncertainty range that is shown in Fig. S11.

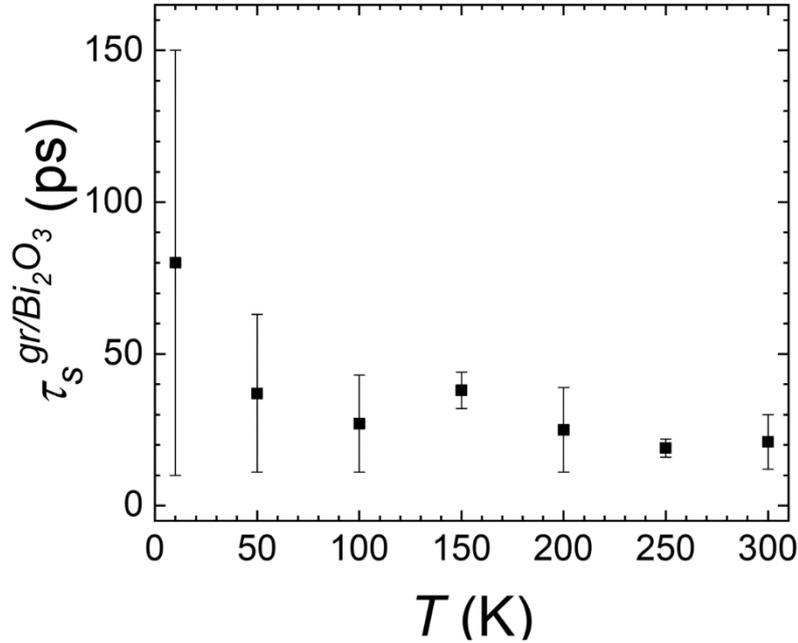

**Figure S12.** Temperature dependence of $\tau_s^{gr/Bi_2O_3}$ obtained following the analysis shown in Figure S11.

## S8. Determination of the spin Hall angle in the graphene/Bi$_2$O$_3$ region

From the analysis described in Section S7, we have obtained the relevant spin transport parameters in the channel and spin polarization of the Co contacts. The next step is to determine the spin Hall angle from $R_{SCC}$.

For this purpose, we need the spin accumulation $\mu_z$ in region $III$ of the five-region model described in section S.7.2.3., which allows us to determine the spin-to-charge conversion signal $R_{SCC}$. In particular, the transverse charge current generated by the inverse spin Hall effect is proportional to the average $I_s^z$ across region $III$, which we obtain using the following expression:

$$\overline{I_s^z} = \frac{1}{W_{cr}} \int_L^{L+W_{cr}} I_s^z(x) dx$$

and the spin-to-charge conversion signal is given by:

$$R_{SCC}^{prec} = \theta^{gr/Bi_2O_3} R_{sq}^{gr/Bi_2O_3} \overline{I_s^z}$$

with $\theta^{gr/Bi_2O_3}$ the spin Hall angle and $R_{sq}^{gr/Bi_2O_3}$ the square resistance of the $Bi_2O_3$-covered region.

As described in section (contact pulling), the application of $B_x$ also leads to the pulling of the contact magnetization. Hence the measured data in the parallel and antiparallel configurations are given by

$$R_{NL}^{\uparrow(\downarrow)} = +(-)R_{SCC}^{prec} \cos(\theta_M) + R_{SCC}^{\parallel} \sin(\theta_M)$$

where $R_{SCC}^{\parallel}$ is the spin-to-charge conversion efficiency along $x$. To extract only the spin Hall signal we subtract both measurements and obtain

$$R_{SCC} = \frac{R_{NL}^{\uparrow} - R_{NL}^{\downarrow}}{2} = R_{SCC}^{prec} \cos(\theta_M). \tag{S4}$$

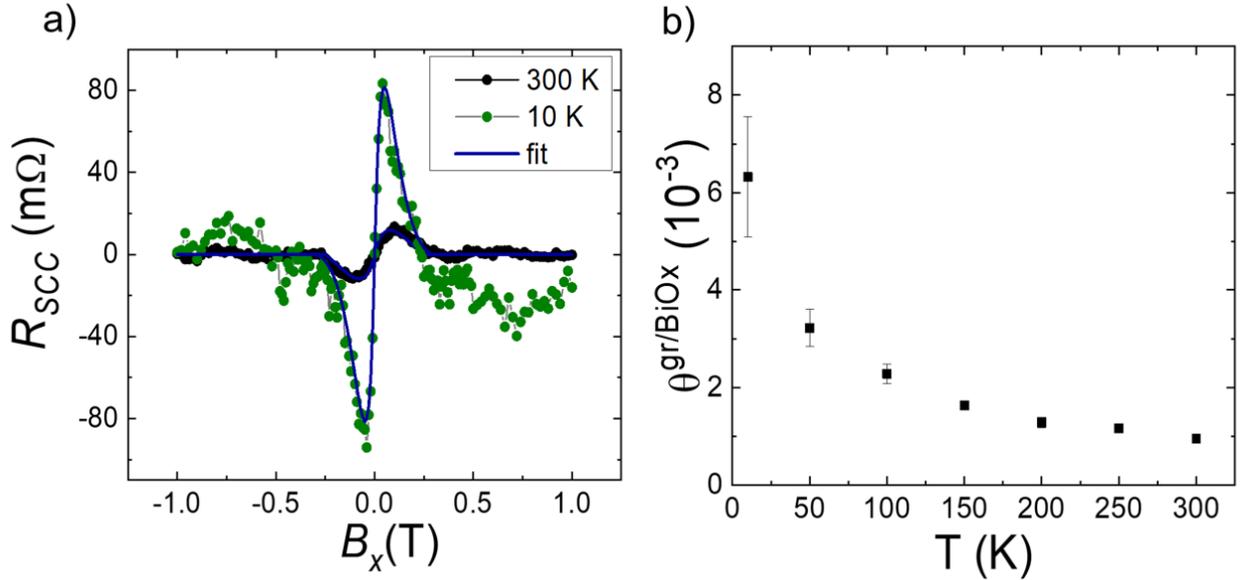

**Figure S13**. (a) Fits to the antisymmetric Hanle curve ($R_{SCC}$ vs $B_x$) at 300 K and 10 K. (b) Extracted spin Hall angle as a function of temperature.

We perform this operation by fitting equation S4 to $R_{SCC}$ obtained experimentally with $\theta^{gr/Bi_2O_3}$ as the only fitting parameter and the contact pulling obtained from the Stoner-Wohlfarth model (Figure S8). The results from that operation are shown in Figure S13 with the corresponding error ranges determined by the uncertainties in $\tau_s^{gr/Bi_2O_3}$ and in the fit to the antisymmetric Hanle data assuming that both uncertainties are not correlated. The relative error ranges are significantly smaller than those of $\tau_s^{gr/Bi_2O_3}$ even though these have been considered. We attribute it to the fact that the spin-to-charge conversion occurs at the whole width of the Bi$_2$O$_3$-covered region and, hence, the spin current entering the left edge dominates the spin-to-charge conversion, decreasing the role of $\lambda_s^{gr/Bi_2O_3}$ on this uncertainty.